%% file: envAwareMIMO_v3_arXiv.tex
\documentclass[conference]{IEEEtran}
\usepackage{amsfonts}

\usepackage{enumerate}

\usepackage{amsmath,amsthm}
\usepackage[noend]{algpseudocode}
\usepackage{float}
\usepackage{color}
\usepackage{makeidx}
\usepackage{bbm}
\usepackage{graphicx}
\usepackage{lipsum}
\usepackage{soul}
\usepackage{tabularx}
\usepackage{dsfont}
\usepackage{balance}
\input{input.tex}
\usepackage{epstopdf}

\newcommand{\sref}[1]{{Section}~\ref{#1}}

\DeclareMathOperator*{\argmax}{arg\,max}

\newcommand{\subto}{\operatorname{s.t.}}

\begin{document}
\title{Learning Beam Codebooks with Neural Networks: Towards Environment-Aware  mmWave  MIMO}
\author{\IEEEauthorblockN{Yu Zhang, Muhammad Alrabeiah,  and Ahmed Alkhateeb}
	\IEEEauthorblockA{\textit{School of Electrical, Computer, and Energy Engineering} \\ \textit{ Arizona State University}\\ Email: {y.zhang, malrabei, alkhateeb}@asu.edu}
	}

\maketitle

\begin{abstract}
Scaling the number of antennas up is a key characteristic of current and future wireless communication systems. The hardware cost and power consumption, however, motivate large-scale MIMO systems, especially at millimeter wave (mmWave) bands, to rely on analog-only or hybrid analog/digital transceiver architectures. With these architectures, mmWave base stations normally use pre-defined beamforming codebooks for both initial access and data transmissions. Current beam codebooks, however, generally adopt single-lobe narrow beams and scan the entire angular space. This leads to high beam training overhead and loss in the achievable beamforming gains. In this paper, we propose a new machine learning framework for learning beamforming codebooks in hardware-constrained large-scale MIMO systems. More specifically, we develop a neural network architecture that accounts for the hardware constraints and learns beam codebooks that adapt to the  surrounding environment and the user locations. Simulation results highlight the capability of the proposed solution in learning multi-lobe beams and reducing the codebook size, which leads to noticeable gains compared to classical codebook design approaches.
\end{abstract}

\section{Introduction} \label{intro}

Millimeter wave (mmWave) and Terahertz (THz) communication systems are essential components of 5G and beyond \cite{6GAndBeyond,HeathJr2016}. To guarantee sufficient received signal power, these systems deploy large antenna arrays and use narrow beams. Due to the high cost and power consumption of high frequency RF chains, however, mmWave and THz communication systems can not use fully-digital architectures that assign an RF chain per antenna, and normally resort to analog-only or hybrid analog/digital architectures \cite{Alkhateeb2014}. Further, because of the hardware constraints on these large-scale MIMO systems, they typically adopt pre-defined single-lobe beamforming codebooks (such as DFT codebooks \cite{HeathJr2016,Hur2013}) that scan all possible directions for both initial access and data transmission. These codebooks have two key limitations: (i) They incur high beam training overhead by scanning all possible directions even though many of these directions may never be used, and (ii) they normally have single-lobe beams which may not be optimal, especially in non-line-of-sight (NLOS) scenarios.

\textbf{Contribution:} In this paper, we consider hardware-constrained large-scale MIMO systems and propose an artificial neural network based framework for learning environment aware beamforming codebooks. More specifically, we design a machine learning model that can learn how to adapt the patterns of the codebook beams based on the surrounding environment and user distribution. This is done by developing a novel complex-valued neural network architecture in which the weights directly model the beamforming/combining weights. Further, the developed neural network architecture accounts for the hardware constraints (such as phase-only, constant-modulus, and quantized-angle constraints) that are typically applied on the analog-only mmWave/THz transceiver architectures. We evaluate the performance of the proposed solution in both outdoor LOS and indoor NLOS communication scenarios. The simulation results highlight the capability of the proposed solution in learning multi-lobe beams that adapt to the environment geometry and to where the users are. Further, for the adopted NLOS simulation setup, the results show that the learned codebook can exceed the 64-beam DFT codebook with only 16 beams, which can lead to noticeable savings in the beam training overhead.

\textbf{Prior work:} Designing beamforming codebooks has been an important research topic for long time at both academia and industry (see \cite{Love2008} for a good overview).  With large-scale MIMO systems, however, the hardware limitations (especially at mmWave/THz) and the use of analog-only or hybrid transceiver architectures imposed new constraints on the codebook design problems. This motivated the development of new beamforming codebooks \cite{Alkhateeb2014,Hur2013,Mo2019}. These codebooks, however, are generally designed to have single-lobe narrow beams that cover all the angular directions and are not adaptive to the deployment, surrounding environment, and user distributions. This motivated the development of environment and hardware aware codebook learning  strategies, which is the focus of this paper.

\section{System and Channel Models} \label{sec:System}

In this paper, we consider a system model where a mmWave base station (BS), equipped with $M$ antennas, is communicating with a single-antenna user. In the uplink transmission, if the user transmits a symbol $x\in\mathbb{C}$ to the BS, with the power constraint $\mathbb{E}\left[|x|^2\right]=P$, the received signal at the BS after combining can be expressed as
\begin{equation}\label{sys}
y = {\mathbf w}^H{\mathbf h}x + {\mathbf w}^H{\mathbf n},
\end{equation}
where ${\mathbf h}\in\mathbb{C}^{M\times 1}$ is the uplink channel vector between the mobile user and the BS, ${\bf w}$ is the receive combining vector at the BS and ${\mathbf n}\sim\mathcal{N}_\mathbb{C}\left(0, \sigma_n^2 {\bf I}\right)$ is the receive noise vector at the BS. To reduce the cost and power consumption, the BS is assumed to employ analog-only transceiver architecture \cite{Alkhateeb2014d}, where the beamforming/combining processing is implemented using analog phase shifters. To account for that, each element $w_m$ in the combining vector, $m =1, 2 , ..., M$, is assumed to have the structure $w_m=\frac{1}{\sqrt{M}} e^{j\theta_m}$.

We adopt a general geometric channel model for ${\mathbf h}$ \cite{Alrabeiah2019}. Assume that the signal propagation between the mobile user and the BS consists of $L$ paths. Each path $\ell$ has a complex gain $\alpha_\ell$ and an angle of arrival $\phi_\ell$. Thus, the channel can be written as
\begin{equation}\label{channel}
  {\mathbf h}=\sum\limits_{\ell=1}^{L}\alpha_\ell{\mathbf a}(\phi_\ell),
\end{equation}
where ${\bf a}(\phi_\ell)$ is the array response vector of the BS. The definition of ${\bf a}(\phi_\ell)$ depends on the array geometry.

\section{Problem Definition} \label{sec:Prob}

In this paper, we consider hardware-constrained large-scale MIMO systems and investigate the design of beamforming codebooks that are adaptive to the deployment scenario, environment geometry, user distributions, etc.---which we refer to as environment-aware beamforming codebooks.

Given the system and channel models in \sref{sec:System}, and due to the hardware constraints on the BS transceiver, the BSs normally adopt pre-defined beamforming codebooks. If $\boldsymbol{\mathcal{W}}=\left\{\bw_1, ...., \bw_N\right\}$ denotes the BS beamforming codebook,  then the maximum beamforming/combining gain of user $u$ can be written as
\begin{equation}\label{single_snr}
\gamma_u=  \max\limits_{\bw_n \in \boldsymbol{\mathcal{W}}}\left| {\bf w}_n^H{\bf h}_u \right|^2.
\end{equation}

Our objective in this paper is then to design the beamforming codebook $\boldsymbol{\mathcal{W}}$ to maximize the maximum  beamforming gain $\gamma_u$ averaged over the set of users served by this BS in the surrounding environment.  Let $\mathcal{H}$ represent the set of channel vectors for the candidate users, then our objective can be written as
\begin{align}\label{Prob-0}
 \boldsymbol{\mathcal{W}}_{\mathsf{opt}} = \argmax\limits_{\boldsymbol{\mathcal{W}}} & \hspace{2pt}  \frac{1}{|\mathcal{H}|}\sum_{{\bf h}_u\in\mathcal{H}} \left[  \max\limits_{\bw_n \in \boldsymbol{\mathcal{W}}}\left| {\bf w}_n^H{\bf h}_u \right|^2 \right], \\
 &\hspace{-30pt} \subto  \hspace{2pt}  \left|w_{m n}\right| = \frac{1}{\sqrt{M}}, ~ \forall m\in [M], ~ \forall n\in [N], \label{unit}
\end{align}
where  $| \boldsymbol{\mathcal{W}}|=N$ is the codebook size, and $[N]$ is the shorthand for representing set $\{1,2,\dots,N\}$. The constraint \eqref{unit} is imposed to uphold the phase-shifters constraint.

Constructing efficient beamforming codebooks has attracted significant interest from both academia and industry for the last two decades (see \cite{Love2008} for a good overview). With large-scale MIMO systems, however, the hardware limitations, such as the constant modulus captured by \eqref{unit} and the quantized phase shifters, add hard and non-convex constraints to the beam codebook optimization problems and make them difficult to solve \cite{Alkhateeb2014}.  In this paper, we make an initial step into leveraging the learning and optimization capabilities of neural networks to learn beam codebooks that are adaptive to the environment geometry, user distributions, etc., and that respect the hardware constraints on the large-scale MIMO systems. We believe that this direction of using neural networks to develop  environment and hardware aware beam codebooks could lead in the future extensions to interesting solutions, especially when considering more complex system models such as interference-limited and multi-user MIMO.

\section{Proposed Machine Learning Solution} \label{sec:Sol}

In this section, we present our proposed beam codebook learning solution for hardware-constrained large-scale MIMO systems. First, we explain in \sref{sup_arch} the proposed neural network architecture  and its relation to the optimization problem \eqref{Prob-0}-\eqref{unit}, before going into the details of how a codebook is learned in \sref{sec:CB_learn}.

\subsection{Model Architecture} \label{sup_arch}

\begin{figure}[t]
	\centering
	\includegraphics[width=1\linewidth]{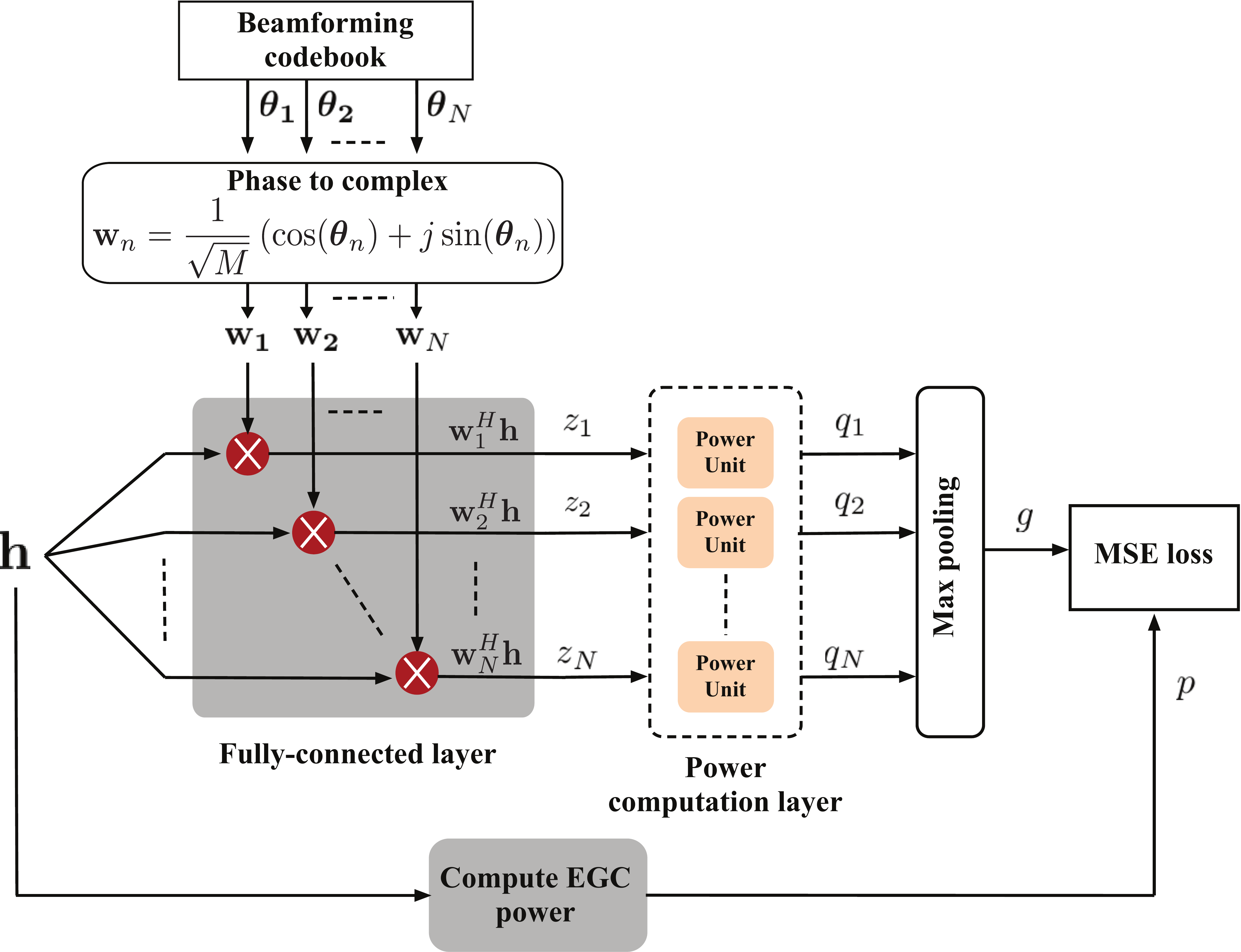}
	\caption{This schematic shows the overall architecture of the neural network used to learn beamforming codebooks. It highlights the network architecture and the auxiliary components, equal-gain-combining and MSE-loss units, used during the training process. It also gives a slightly deeper dive into the inner-workings of the cornerstone of this architecture, the complex-valued fully-connected layer.}
	\label{Arch}
\end{figure}

This proposed neural network architecture consists of three main components, as depicted in \figref{Arch}. Those components are the complex-valued fully-connected layer, the power-computation layer and the max-pooling layer. A forward pass through these three layers is equivalent to evaluating the objective function of \eqref{Prob-0} over a single user's channel $\mathbf h$.

\subsubsection{Complex-valued fully-connected layer}

The first layer consists of $N$ neurons that are capable of performing complex-valued multiplication and summation. Each neuron, as shown in \figref{Arch}, learns one beamforming vector and performs inner product with the input channel vector. Formally, this is described by the following matrix multiplication
\begin{align}\label{fc_1}
  \mathbf z &= \mathbf W ^H \mathbf h,
\end{align}
where $\mathbf W = [ \mathbf w_1,\dots,\mathbf w_N]\in \mathbb C^{M\times N} $ is the beamforming codebook matrix,  $\mathbf h\in \mathbb C^{M\times 1}$ is a user's channel vector, and $\mathbf z \in \mathbb C^{N\times 1}$ is the vector of the combined received signal. \eqref{fc_1} could be re-written in the following equivalent real-valued block matrix form
\begin{align}\label{fc_2}
	\left[ \begin{array}{l}
		\mathbf z^r \\
		\mathbf z^{im}
	\end{array}\right]
	&= \left[\begin{array}{cc}
  	\mathbf W^r & \mathbf {-W}^{im} \\
  	\mathbf W^{im} & \mathbf W^{r}
  \end{array} \right]^T
  \left[ \begin{array}{l}
  	\mathbf h^r \\
  	\mathbf h^{im}
  \end{array}\right],
\end{align}
where $\mathbf z^r$, $\mathbf z^{im}$ are the real and imaginary parts of $\mathbf z$, $\mathbf W^r$, $\mathbf W^{im}$ are matrices containing the real and imaginary components of the elements of $\mathbf W$, and similarly, $\mathbf h^r$, $\mathbf h^{im}$ are the real and imaginary components of the channel vector $\mathbf h$.

Contrary to the norm in designing neural networks, the elements of the beamforming matrix $\mathbf W$ are not the weights of the fully-connected layer. Instead, they are derived from the actual neural network weights, which are the phases of the phased arrays. This is done through an embedded layer of phase-to-complex operations, as shown in \figref{Arch}. This layer transforms the phased arrays into unit-magnitude complex vectors by applying elements-wise $\cos$ and $\sin$ operations and scaling them by $1/\sqrt{M}$ as follows
\begin{align}
  \mathbf W &= \frac{1}{\sqrt{M}}\left( \cos\left( \mathbf \Theta\right) + j
  \sin\left( \mathbf \Theta \right) \right),
\end{align}
where $\mathbf \Theta = \left[ \boldsymbol{\theta}_1, \dots, \boldsymbol{\theta}_{N} \right]$ is an $M \times N$ matrix that includes the phases of the beamforming codebook, and $\boldsymbol{\theta}_n = \left[\theta_{1n}, \dots, \theta_{Mn}\right]^T$, $\forall n \in \{1,\dots, N\}$ is a single phase vector. The use of this embedded layer is the network's way of learning beamforming vectors that respect the phase shifter constraint.

\subsubsection{Power-computation layer}

The output of the complex-valued fully-connected layer is then fed into the power-computation layer. It performs element-wise absolute square operation and outputs a real-valued vector ${\bf q}$ given by
\begin{equation}\label{powOut}
  {\bf q} = \left[q_1, q_2, \dots, q_N\right]^T = \left[ |z_1|^2, |z_2|^2, \dots, |z_N|^2 \right]^T,
\end{equation}
which contains the received power of each beamforming vector in the codebook.

\subsubsection{Max-pooling layer}

The power of the best beamforming vector is finally found by the last layer, the max-pooling layer. It performs the following element-wise $\max$ operation over the elements of $\mathbf q$, that is
\begin{equation}\label{maxOut}
  g = \max\left\{q_1, q_2, \dots, q_N\right\},
\end{equation}
where $g$ is the power of the best beamforming vector. This value is used to assess the quality of the codebook by comparing it to a desired receive power value. In the following subsection, we describe the details of the desired value and how the quality is assessed.

\subsection{Learning Codebooks} \label{sec:CB_learn}

With the learning architecture in mind, it is time to delve into the details of how a codebook is learned. Our proposed solution follows a supervised learning approach. In such approach, a machine learning model is trained using pairs of inputs and the desired responses, which constitute the  dataset.

\subsubsection{Dataset Description}

The inputs to the model are the users' channels as they are the communication quantity that drives the beamforming design process. For labels, there are many possible desired responses that could be used for training, and the choice between them should be made based on what the model needs to learn. In this paper, we adopt the equal gain combining (EGC) as the model label. This choice is based on the facts that EGC respects the phase shifters constraint and it is the beamforming strategy that achieves optimal beamforming gain  when there are no restrictions on the codebook size or the quantized phase shifters. The EGC beamforming vector is constructed by using the phase component of every user's channel as  $ {\bf f}_\mathrm{EGC} = \frac{1}{\sqrt{M}} e^{\angle \bh}$, where $\angle \bh$ stands for the vector of phases of a complex vector $\bh$. Using EGC beamforming vectors, the desired response for each user can be computed as follows
\begin{equation}\label{EGCgain}
  p = \left| {\bf f}_\mathrm{EGC}^H{\bf h} \right|^2 = \left\|{\bf h}\right\|_1^2,
\end{equation}
where $\|\cdot\|_1$ is the $L_1$ norm of a vector.
Putting the users' channels and their EGC gains together provides the training dataset $\mathcal S_{t}$. This is formally given by
\begin{equation}\label{t_set}
  \mathcal S_t = \left\{ (\mathbf h_1, p_1), \dots,(\mathbf h_U, p_U) \right\},
\end{equation}
where $\mathbf h_u \in \mathbb C^{M\times 1}$ and $p_u \in \mathbb R$ are the channel vector and EGC gain of the $u$-th user respectively, and $U=|\mathcal{H}|$ is the total number of collected data pairs.

\begin{figure}[t]
	\centering
	\subfigure[]{\includegraphics[width=0.57\linewidth]{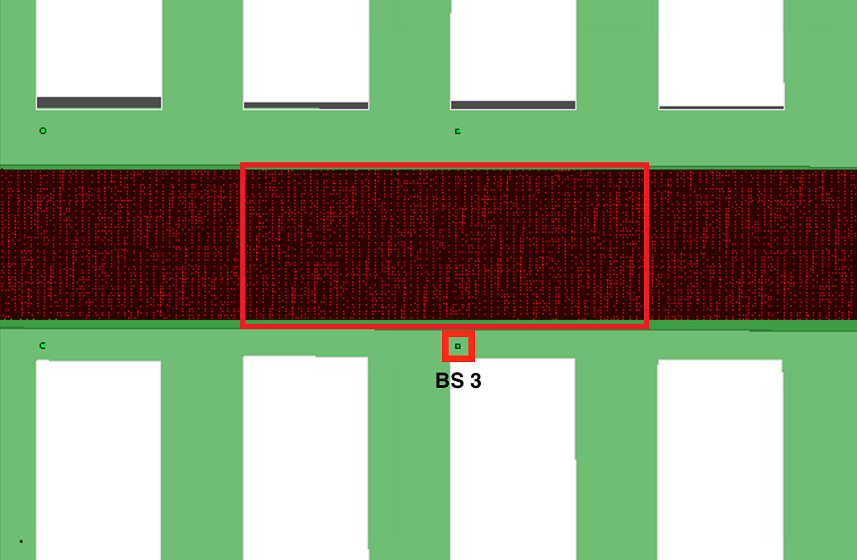}}
	\subfigure[]{\includegraphics[width=0.40\linewidth]{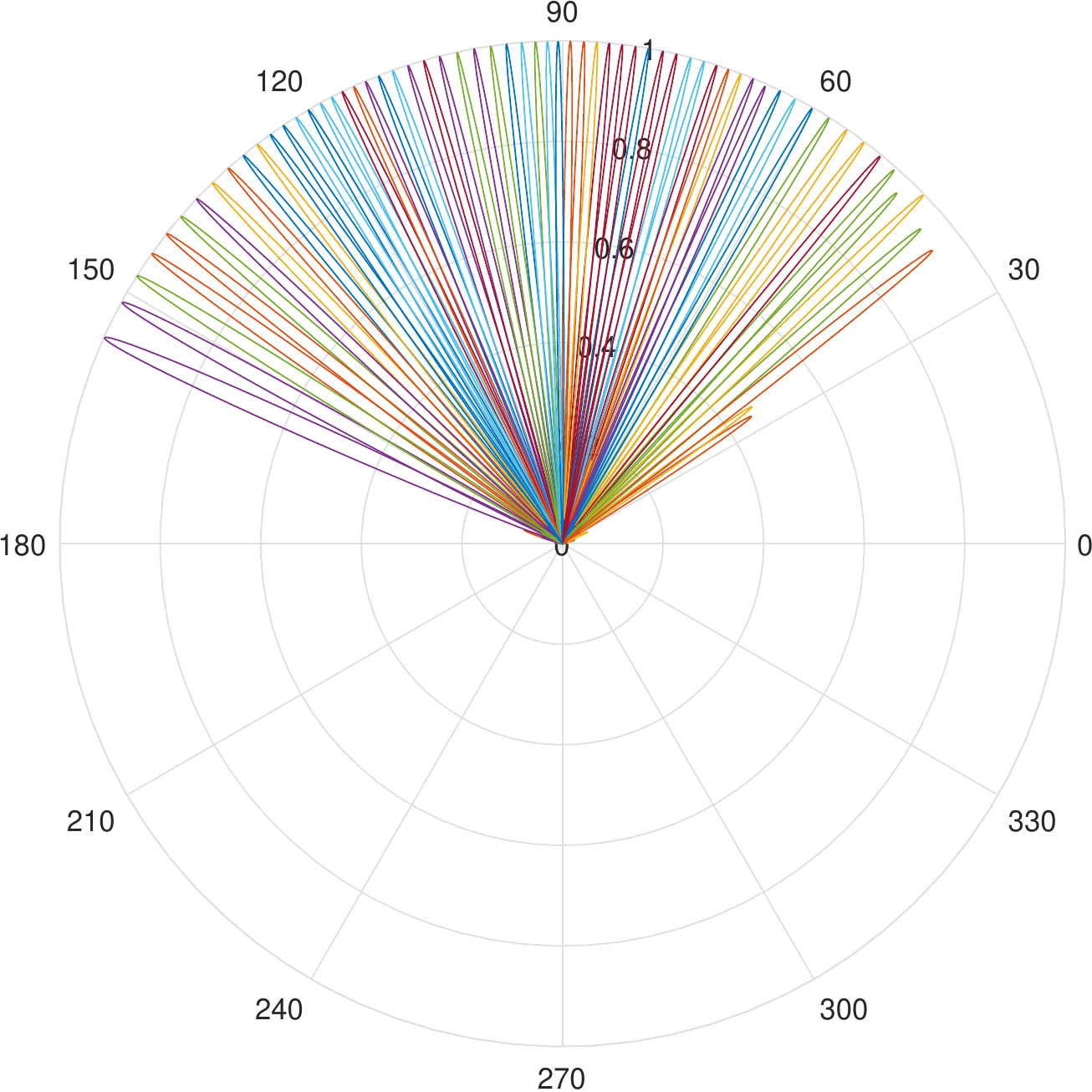}}
	\caption{The demonstration of (a) the adopted LOS scenario and (b) the 64-beams codebook learned for the scenario in (a). It shows how the learned codebook adapt its beams to the user distribution, instead of evenly spreading beams in the whole space like DFT codebook.}
	\label{LOS_beam}
\end{figure}

\subsubsection{Model Training} \label{sup_model_training}

Using the set $\mathcal S_t$, the model is trained by undergoing multiple forward-backward cycles. In each cycle, a \textit{mini-batch} of complex channel vectors and their EGC responses is sampled from the training set. The channels are fed sequentially to the model and a forward pass is performed as describe in \sref{sup_arch}. For each channel vector in the batch, the model combines it with the currently available $N$ beamforming vectors and outputs the power of the best beamforming vector for that channel. The quality of the best combiner is assessed by measuring how close its beamforming gain to that of the equal gain combiner, obtained by \eqref{EGCgain}. A Mean Squared Error (MSE) loss is used as a metric to assess the quality of the codebook over the current mini-batch. Formally, it is defined as
\begin{equation}\label{MSE-loss}
  \mathcal L = \frac{1}{B} \sum_{b=1}^B (g_b - p_b)^2,
\end{equation}
where $g_b$ is the output of the max-pooling layer for the $b$-th data pair in the mini-batch, $p_b$ is the desired value achieved by \eqref{EGCgain} and $B$ is the mini-batch size.

\begin{figure}[t]
	\centering
	\includegraphics[width=0.66\columnwidth]{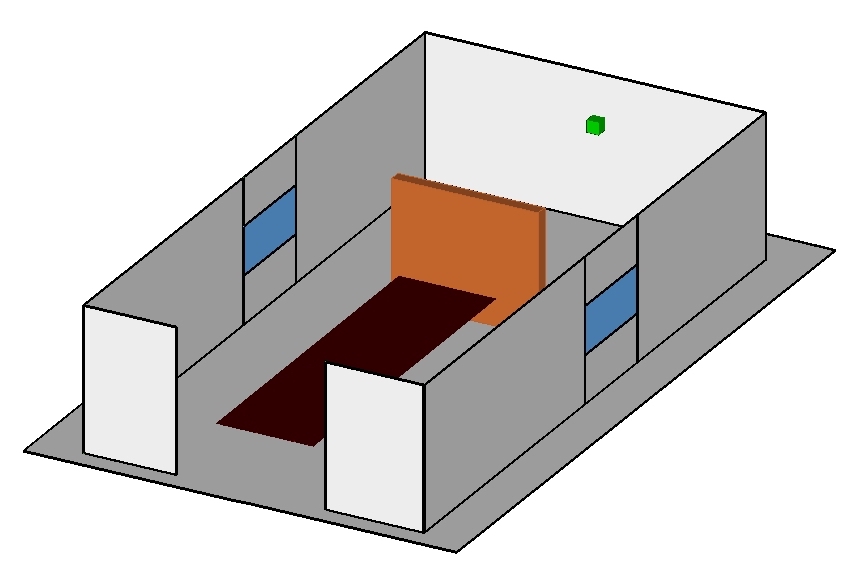}
	\caption{The adopted NLOS scenario, which is chosen to be indoor for the high likelihood of having NLOS connection there. There is no LOS connection between any user in the grid and the antenna array.}
	\label{NLOS_scenario}
\end{figure}

The error signal (derivative of the loss \eqref{MSE-loss} with respect to each phase vector $\boldsymbol \theta_n \in \mathbf{\Theta}$) is back propagated through the model to adjust the phases of the combining vectors \cite{LearningMachines} \cite{EffBackProp}, making up what is usually referred to as the backward pass or \textit{backpropagation}. This is formally expressed by the chain rule of differentiation
\begin{equation}\label{chain_rule}
  \left(\frac{\partial \mathcal L}{\partial \boldsymbol{\theta}_n}\right)^T = \frac{\partial \mathcal L}{\partial g} \cdot \left(\frac{\partial g}{\partial \mathbf q}\right)^T \cdot \frac{\partial \mathbf q}{\partial \mathbf z} \cdot \frac{\partial \mathbf z}{\partial \boldsymbol{\theta}_n}.
\end{equation}
Mathematically, $\frac{\partial \mathcal L}{\partial \boldsymbol{\theta}_n}$ does not exist for two reasons: 1) the factor $\frac{\partial g}{\partial \mathbf q}$ does not exist since the max-pooling function $g$ is not differentiable, and 2) the factor $\frac{\partial \mathbf q}{\partial \mathbf z}$ does not satisfy the Cauchy-Riemann equations \cite{Haslinger2017}, meaning that $\mathbf q$ as a function of complex vector ${\bf z}$ is not complex differentiable (\textit{holomorphic}). However, both issues could be resolved to enable backpropagation. The details of that are discussed in \cite{Trabelsi2017}. Computing the partial derivative of loss with respect to phase vector $\boldsymbol{\theta}_n$ allows the backward pass to modify the codebook $\boldsymbol \Theta$ and make it adaptive to the environment. The update equation generally depends on the solver used to carry out the training, e.g., Stochastic Gradient Descent (SGD) and ADAptive Moment estimation (ADAM) to name two, but in its simplest form, it could be given by
\begin{equation}
  \boldsymbol{\theta}_{n_\mathsf{new}} = \boldsymbol{\theta}_{n_\mathsf{old}} - \eta \cdot \frac{\partial \mathcal L}{\partial \boldsymbol{\theta}_n},
\end{equation}
where $\eta$ is the optimization step size, commonly known as the learning rate in machine learning, and $\boldsymbol{\theta}_{n_\mathsf{new}}$, $\boldsymbol{\theta}_{n_\mathsf{old}}$ are the new and current $n$-th phase vector of the codebook respectively.

Through the forward-backward cycles, the model learns to adapt its beamforming vectors according to the channels it experiences in the mini-batch. As mentioned in Section \ref{sup_arch}, a forward pass is equivalent to evaluating the objective function in \eqref{Prob-0}. In a similar spirit, the backward pass through the architecture independently \textit{optimizes} the beamforming vectors based on the \textbf{group of channels} that they receive best. It turns out that such group of channels that are best received by the same beamforming vector contribute only to the update of that beamforming vector and do not affect the others. This mainly comes as a result of the masking role the max-pooling layer plays in the backward pass.

\section{Experimental Setup and Model Training} \label{sec:Exp}

In this section, we describe the adopted scenarios, datasets and the machine learning parameters used in our simulation.

\begin{figure}[t]
	\centering
	\subfigure[]{\includegraphics[width=0.40\linewidth]{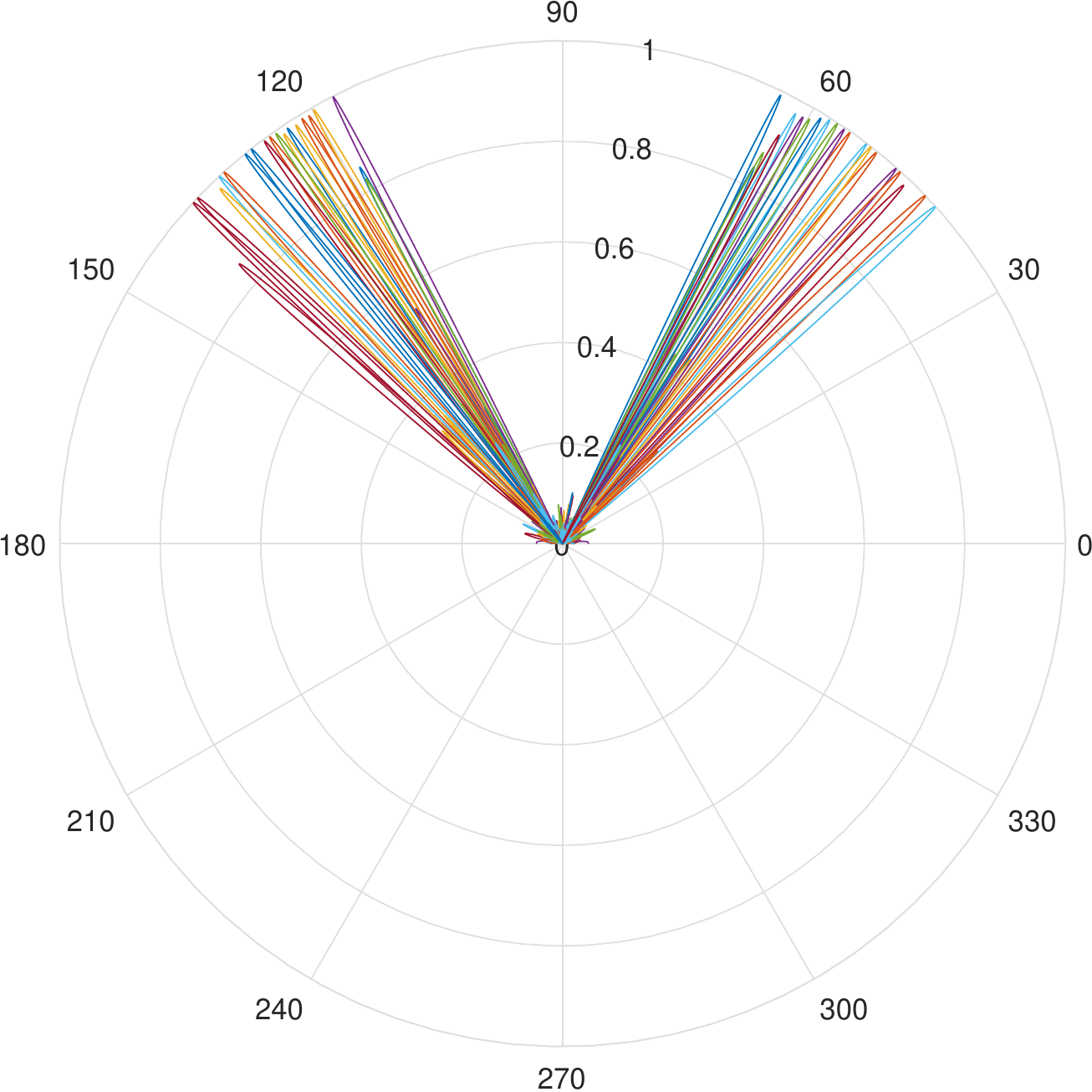}}
	\subfigure[]{\includegraphics[width=0.40\linewidth]{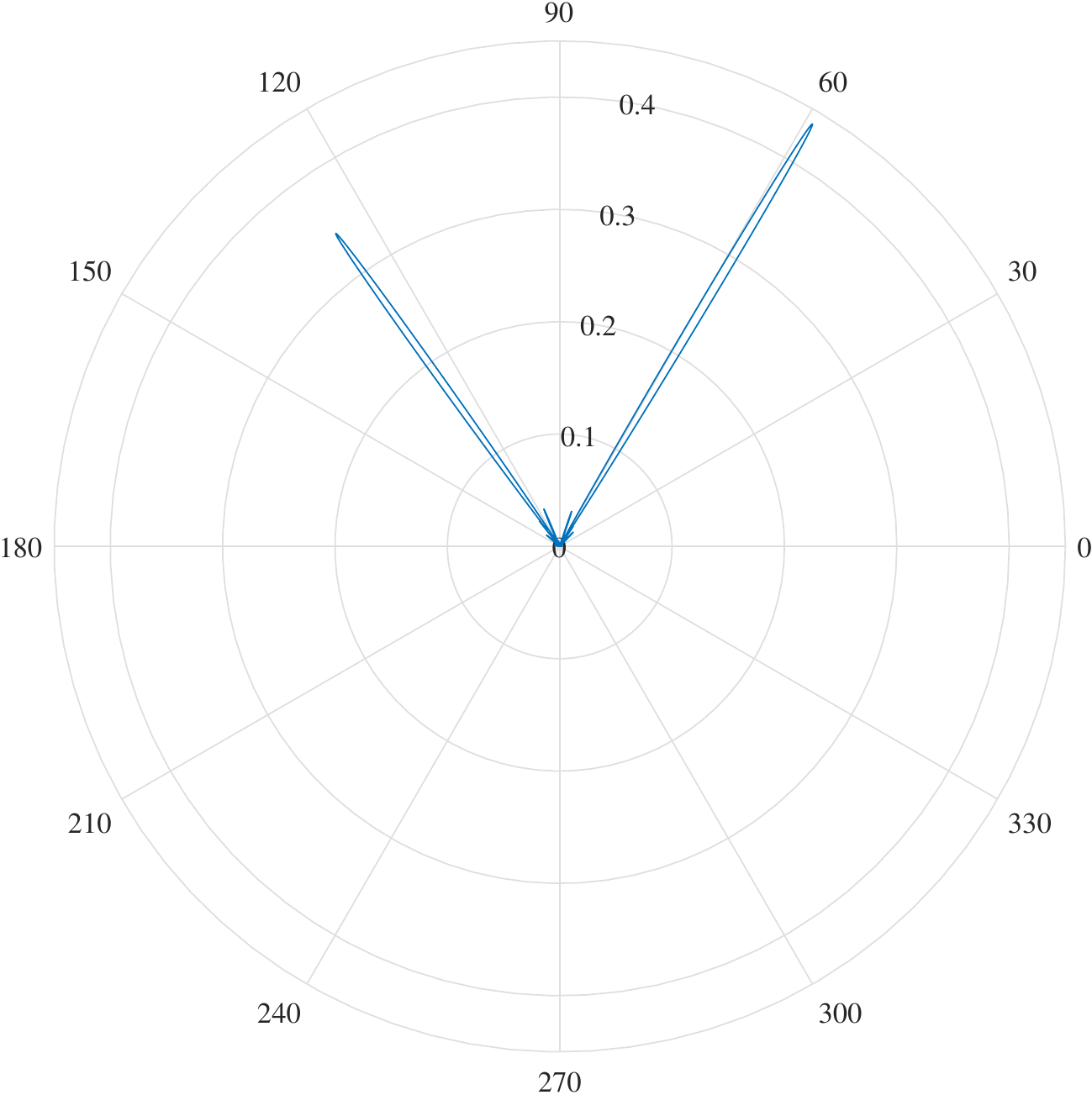}}
	\caption{The learned codebook for the scenario in \figref{NLOS_scenario}. (a) shows all the 64 beams in the codebook, and (b) demonstrates one of them with multi-lobes adapted for the different NLOS paths.}
	\label{NLOS_beam}
\end{figure}

\subsection{Scenario and Dataset} \label{subsec:Sc}

The performance of the proposed solution is evaluated on two different communication scenarios. The first is an outdoor scenario which addresses the situation when all users experience LOS connection with the base station, named LOS scenario. The second, on the other hand, is an indoor scenario and addresses the case when none of the users has a LOS connection, which is referred to as NLOS  scenario.
Both scenarios are for an operating frequency of 28GHz, and both are part of the scenarios in the DeepMIMO dataset \cite{DeepMIMO} . Using the data generation script of DeepMIMO, two sets of channel data are generated, one for each scenario.
Table \ref{param} shows the parameters of channel data generation.

\begin{table}[t]
	\caption{The parameters of the DeepMIMO dataset}
	\centering
	\begin{tabular}{|c | c | c|}
		\hline
		Name of scenario & O1\textunderscore28 & I2\textunderscore28B \\
		\hline
		Active BS & 3 & 1 \\
		\hline
		Active users & 700 to 1300 & 1 to 700 \\
		\hline
		Number of antennas (x, y, z)  & (1, 64, 1) & (64, 1, 1) \\
		\hline
		System BW & 0.2 GHz & 0.2 GHz\\
		\hline
		Antenna spacing & 0.5 & 0.5 \\
		\hline
		Number of OFDM sub-carriers & 1 & 1 \\
		\hline
		OFDM sampling factor & 1 & 1 \\
		\hline
		OFDM limit & 1 & 1 \\
		\hline
	\end{tabular}
	\label{param}
\end{table}

\subsection{Model Training and Testing}

The model is trained and tested by using both datasets described in \sref{subsec:Sc}. Prior to any training session, the channels of each dataset is pre-processed to improve the training experience \cite{EffBackProp}, which is a very common practice in machine learning. The pre-processing of choice for these experiments is data sample normalization. As in \cite{Alrabeiah2019, Sub6PredMmWave, zhang2019deep, Li2019,Alkhateeb2018a}, the channel normalization using the maximum absolute value in the training dataset helps the network undergo a stable and efficient training. Formally, the normalization factor is found as follows
\begin{equation}
\Delta = \underset{\mathbf h_u\in S_t}{\max} \ \left|[\bh_u]_m \right|^2
\end{equation}
where $[\bh_u]_m$ is the $m$-th element in the channel vector of the $\bh_u$. Using the normalized channels, the model is trained on portion of the samples of the dataset and tested on the rest. For brevity, the data split percentage between training and testing along with other training hyper-parameters can be found at \cite{myGithub}.

\section{Simulation Results} \label{sec:Simu}

In this section, we evaluate the performance of our proposed machine learning based beamforming codebook design solution on the scenarios described in \sref{subsec:Sc}. The numerical results show that our proposed model can achieve performance that is comparable to EGC and adapt to different user distributions.

\begin{figure}[t]
	\centering
	\includegraphics[width=1\columnwidth]{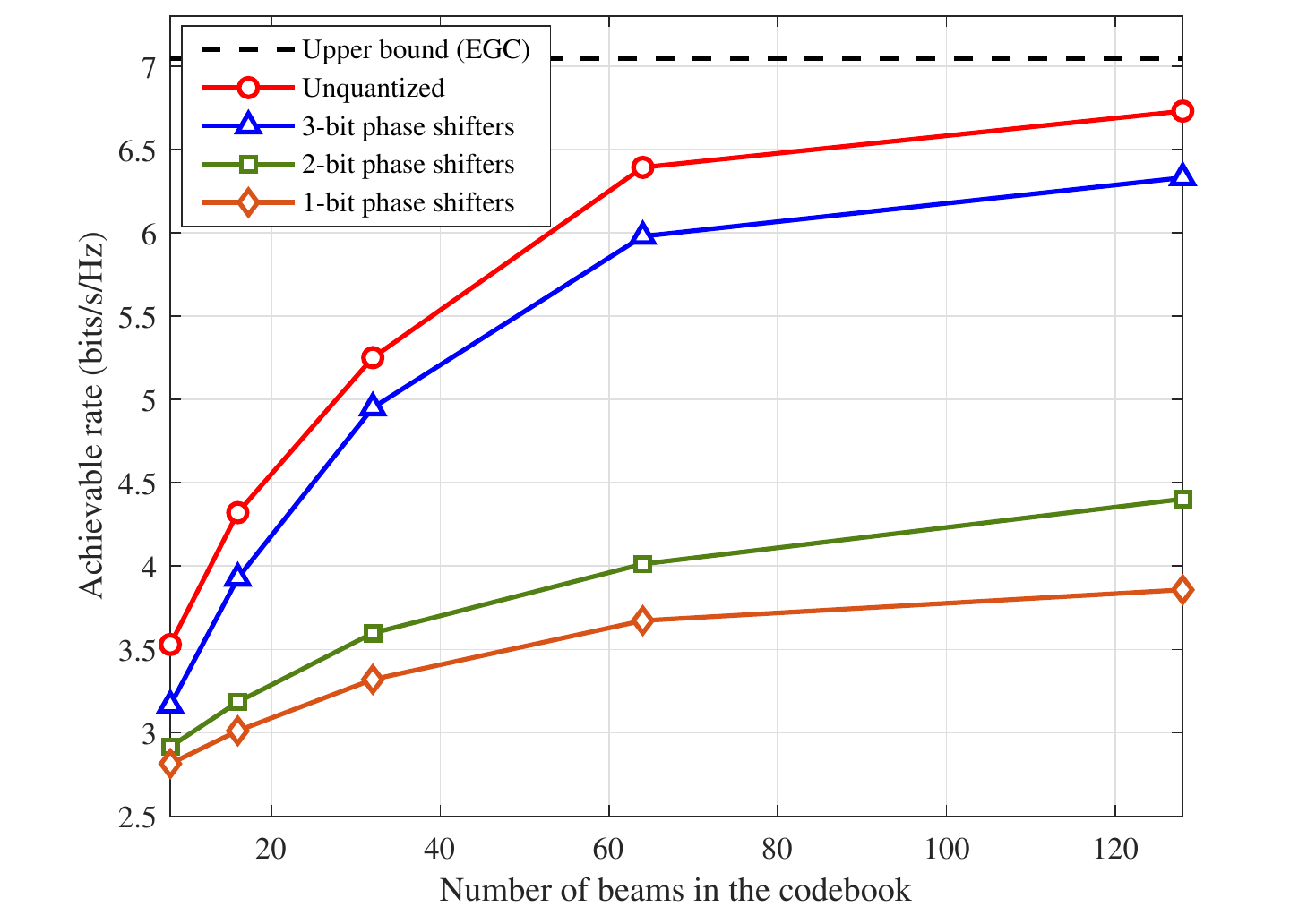}
	\caption{Achievable rate versus number of beams in the codebook in an outdoor LOS scenario under the receive SNR of 5 dB with different bits of quantization.}
	\label{sup_noise_5dB_LOS}
\end{figure}

\figref{sup_noise_5dB_LOS} shows the evaluation results of the proposed solution in a LOS scenario. With 64 beams, the learned codebook reaches more than $90\%$ of the upper bound, which is achieved by the EGC where the number of beams in the codebook is equal to the number of users (and assuming unquantized phase shifters).  Further, with 128 beams, the developed approach achieves more than $95\%$ of the upper bound. Besides, \figref{sup_noise_5dB_LOS} also exhibits the performance of the codebook with quantized phase shifters. It shows that with only 3-bit phase shifters, the learned codebook can still achieve more than $85\%$ of the upper bound with 64 beams and $90\%$ of the upper bound with 128 beams. This is very important and interesting for cases where the resolution of the analog phase shifters is limited.

\begin{figure}[t]
	\centering
	\includegraphics[width=1\columnwidth]{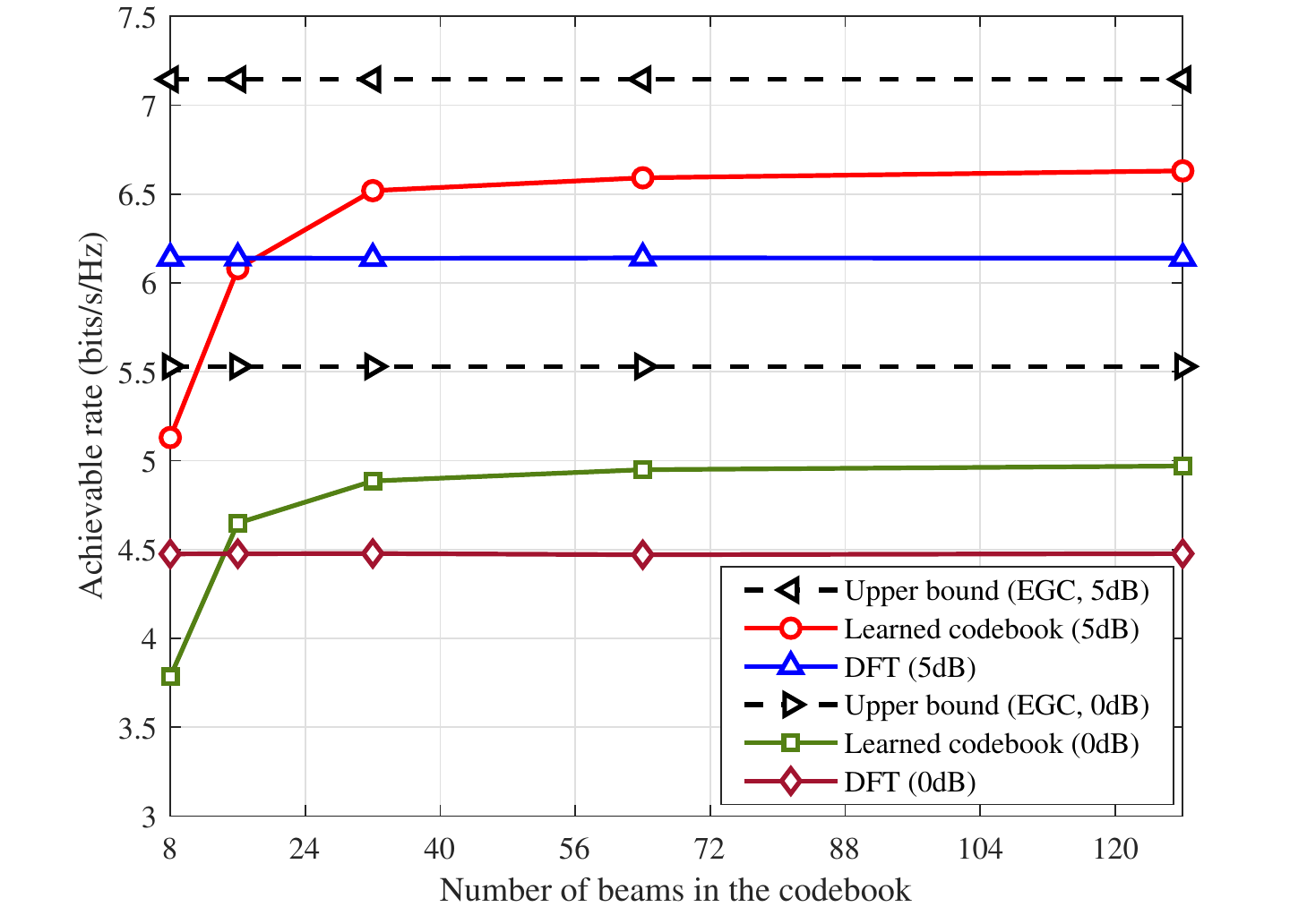}
	\caption{The achievable rate versus number of beams in the codebook in an indoor NLOS scenario under the receive SNR of 0 and 5 dB.}
	\label{sup_noise_0dB_5dB_NLOS}
\end{figure}

\figref{sup_noise_0dB_5dB_NLOS} shows the evaluation result of the proposed solution in a NLOS scenario. This is more interesting and challenging than the LOS case. In a NLOS scenario, each user may have multiple rays reflecting off some scatters and reaching the base station. As depicted in \figref{NLOS_beam}, the learned codebook with the developed approach is capable of learning multi-lobe beams that capture the dominant rays and gather more energy compared to classical DFT codebooks. \figref{sup_noise_0dB_5dB_NLOS} depicts the performance of the proposed solution in this NLOS environment. \textbf{The learned codebook has almost the same performance or even outperforms a 64-beams DFT codebook with only 16 beams, which is one fourth in size}. With 64 beams, the learned codebook achieves almost $90\%$ of the upper bound and noticeable gain compared to the 64-beam DFT codebook.

\section{Acknowledgment}
This work is supported by the National Science Foundation under Grant No. 1923676.

\section{Conclusions and Discussions} \label{sec:Con}

In this paper, we developed a machine learning framework for learning environment and hardware aware beamforming codebooks for large-scale MIMO systems. In the developed solution, the neural networks incorporate the hardware constraints and learn how to adapt the beam shapes based on the surrounding environment and user distribution. Simulation results confirmed the capability of the proposed solution in learning beam codebooks that are optimized in the size and  beam patterns. This leads to promising gains in terms of the achievable rates compared to classical DFT codebooks. In the future, it is interesting to extend the proposed machine learning framework to address more complicated scenarios such as multi-stream and multi-user communications.

\balance

\end{document}

%% file: input.tex
\usepackage{amsfonts}
\usepackage{times}
\usepackage{graphicx}
\usepackage{latexsym}
\usepackage{dsfont}
\usepackage{amssymb}
\usepackage{amsmath}
\usepackage{cite}
\usepackage{verbatim}
\usepackage{subfigure}

\newcommand{\figref}[1]{{Fig.}~\ref{#1}}

\def\bb0{{\mathbb{0}}}

\def\bb{{\mathbf{b}}}

\def\bh{{\mathbf{h}}}

\def\bw{{\mathbf{w}}}

\def\b0{{\mathbf{0}}}

\def\sf0{{\mathsf{0}}}









%% file: envAwareMIMO_v3_arXiv.bbl
\begin{thebibliography}{10}
	\providecommand{\url}[1]{#1}
	\csname url@samestyle\endcsname
	\providecommand{\newblock}{\relax}
	\providecommand{\bibinfo}[2]{#2}
	\providecommand{\BIBentrySTDinterwordspacing}{\spaceskip=0pt\relax}
	\providecommand{\BIBentryALTinterwordstretchfactor}{4}
	\providecommand{\BIBentryALTinterwordspacing}{\spaceskip=\fontdimen2\font plus
		\BIBentryALTinterwordstretchfactor\fontdimen3\font minus
		\fontdimen4\font\relax}
	\providecommand{\BIBforeignlanguage}[2]{{%
			\expandafter\ifx\csname l@#1\endcsname\relax
			\typeout{** WARNING: IEEEtran.bst: No hyphenation pattern has been}%
			\typeout{** loaded for the language `#1'. Using the pattern for}%
			\typeout{** the default language instead.}%
			\else
			\language=\csname l@#1\endcsname
			\fi
			#2}}
	\providecommand{\BIBdecl}{\relax}
	\BIBdecl
	
	\bibitem{6GAndBeyond}
	T.~S. {Rappaport}, Y.~{Xing}, O.~{Kanhere}, S.~{Ju}, A.~{Madanayake},
	S.~{Mandal}, A.~{Alkhateeb}, and G.~C. {Trichopoulos}, ``Wireless
	communications and applications above 100 ghz: Opportunities and challenges
	for 6g and beyond,'' \emph{IEEE Access}, vol.~7, pp. 78\,729--78\,757, 2019.
	
	\bibitem{HeathJr2016}
	R.~W. Heath, N.~Gonzlez-Prelcic, S.~Rangan, W.~Roh, and A.~M. Sayeed, ``An
	overview of signal processing techniques for millimeter wave {MIMO}
	systems,'' \emph{IEEE Journal of Selected Topics in Signal Processing},
	vol.~10, no.~3, pp. 436--453, April 2016.
	
	\bibitem{Alkhateeb2014}
	A.~Alkhateeb, O.~El~Ayach, G.~Leus, and R.~Heath, ``Channel estimation and
	hybrid precoding for millimeter wave cellular systems,'' \emph{IEEE Journal
		of Selected Topics in Signal Processing}, vol.~8, no.~5, pp. 831--846, Oct.
	2014.
	
	\bibitem{Hur2013}
	S.~Hur, T.~Kim, D.~Love, J.~Krogmeier, T.~Thomas, and A.~Ghosh, ``Millimeter
	wave beamforming for wireless backhaul and access in small cell networks,''
	\emph{IEEE Transactions on Communications}, vol.~61, no.~10, pp. 4391--4403,
	Oct. 2013.
	
	\bibitem{Love2008}
	D.~Love, R.~Heath, V.~Lau, D.~Gesbert, B.~Rao, and M.~Andrews, ``An overview of
	limited feedback in wireless communication systems,'' \emph{IEEE Journal on
		Selected Areas in Commun.}, vol.~26, no.~8, pp. 1341--1365, Oct. 2008.
	
	\bibitem{Mo2019}
	J.~{Mo}, B.~L. {Ng}, S.~{Chang}, P.~{Huang}, M.~N. {Kulkarni}, A.~{Alammouri},
	J.~C. {Zhang}, J.~{Lee}, and W.~{Choi}, ``Beam codebook design for 5g mmwave
	terminals,'' \emph{IEEE Access}, vol.~7, pp. 98\,387--98\,404, 2019.
	
	\bibitem{Alkhateeb2014d}
	A.~Alkhateeb, J.~Mo, N.~Gonzalez-Prelcic, and R.~Heath, ``{MIMO} precoding and
	combining solutions for millimeter-wave systems,'' \emph{IEEE Communications
		Magazine,}, vol.~52, no.~12, pp. 122--131, Dec. 2014.
	
	\bibitem{Alrabeiah2019}
	M.~Alrabeiah and A.~Alkhateeb, ``Deep learning for mmwave beam and blockage
	prediction using sub-6ghz channels,'' 2019.
	
	\bibitem{LearningMachines}
	S.~S. Haykin \emph{et~al.}, \emph{Neural networks and learning machines/Simon
		Haykin.}\hskip 1em plus 0.5em minus 0.4em\relax New York: Prentice Hall,,
	2009.
	
	\bibitem{EffBackProp}
	\BIBentryALTinterwordspacing
	Y.~LeCun, L.~Bottou, G.~B. Orr, and K.-R. M\"{u}ller, ``Efficient backprop,''
	in \emph{Neural Networks: Tricks of the Trade}\hskip 1em plus 0.5em minus 0.4em\relax London, UK:
	Springer-Verlag, 1998, pp. 9--50. [Online]. Available:
	\url{http://dl.acm.org/citation.cfm?id=645754.668382}
	\BIBentrySTDinterwordspacing
	
	\bibitem{Haslinger2017}
	F.~Haslinger, \emph{Complex Analysis: A Functional Analytic Approach}.\hskip
	1em plus 0.5em minus 0.4em\relax Walter de Gruyter GmbH \& Co KG, 2017.
	
	\bibitem{Trabelsi2017}
	\BIBentryALTinterwordspacing
	C.~Trabelsi, O.~Bilaniuk, D.~Serdyuk, S.~Subramanian, J.~F. Santos, S.~Mehri,
	N.~Rostamzadeh, Y.~Bengio, and C.~J. Pal, ``Deep complex networks,''
	\emph{CoRR}, vol. abs/1705.09792, 2017. [Online]. Available:
	\url{http://arxiv.org/abs/1705.09792}
	\BIBentrySTDinterwordspacing
	
	\bibitem{DeepMIMO}
	A.~Alkhateeb, ``Deep{MIMO}: A generic deep learning dataset for millimeter wave
	and massive {MIMO} applications,'' in \emph{Proc. of Information Theory and
		Applications Workshop (ITA)}, San Diego, CA, Feb 2019, pp. 1--8.
	
	\bibitem{Sub6PredMmWave}
	M.~{Alrabeiah} and A.~{Alkhateeb}, ``Deep learning for mmwave beam and blockage
	prediction using {Sub-6GHz} channels,'' \emph{submitted to IEEE Transactions
		on Communications, arXiv e-prints}, p. arXiv:1910.02900, Oct 2019.
	
	\bibitem{zhang2019deep}
	Y.~Zhang, M.~Alrabeiah, and A.~Alkhateeb, ``Deep learning for massive {MIMO}
	with 1-bit {ADCs}: When more antennas need fewer pilots,'' \emph{submitted to
		IEEE Wireless Communication Letters, arXiv preprint arXiv:1910.06960}, 2019.
	
	\bibitem{Li2019}
	X.~{Li} and A.~{Alkhateeb}, ``Deep learning for direct hybrid precoding in
	millimeter wave massive {MIMO} systems,'' \emph{in Proc. of Asilomar s, arXiv
		e-prints}, p. arXiv:1905.13212, May 2019.
	
	\bibitem{Alkhateeb2018a}
	A.~Alkhateeb, I.~Beltagy, and S.~Alex, ``Machine learning for reliable mmwave
	systems: Blockage prediction and proactive handoff,'' in \emph{IEEE
		GlobalSIP, arXiv preprint arXiv:1807.02723}, 2018.
	
	\bibitem{myGithub}
	\BIBentryALTinterwordspacing
	[Online]. Available: \url{https://github.com/malrabeiah/learningCB}
	\BIBentrySTDinterwordspacing
	
\end{thebibliography}
